\documentclass[pra,twocolumn,superscriptaddress]{revtex4}
\usepackage{blindtext}
\usepackage{amsfonts}
\usepackage{amsmath,amssymb}
\usepackage{graphicx}
\usepackage{subfigure}
\usepackage{float}
\usepackage{bm}
\usepackage{epstopdf}
\usepackage{url}
\usepackage{hyperref}
\hypersetup{
     colorlinks = true,
     linkcolor = blue,
     anchorcolor = blue,
     citecolor = blue,
     filecolor = blue,
     urlcolor = blue
}

\begin{document}

\title{Quantum protocols in demixed dipolar BEC mixtures in 1D}
\author{C. Madro\~nero and R. Paredes} 
\affiliation{Instituto de F\'{\i}sica, Universidad
Nacional Aut\'onoma de M\'exico, Apartado Postal 20-364, M\'exico D.F. 01000, Mexico.}  
\email{rosario@fisica.unam.mx}

%\pacs{67.85.-d, 03.75.Ss, 03.75.Gg}
\begin{abstract}
Long range dipolar effects in 1D systems either in free or inhomogeneous space are the basis of the state preparation protocol here proposed. Under the presence of an external time-dependent magnetic field, dipole-dipole interactions in the binary ultracold $^{166}$Er -$^{164}$Dy system were tuned from repulsive to attractive to access either, the droplet regime, or the extended one where individual species can be found in mixed or demixed phases. A thorough exploration of weak contact and dipole-dipole effective interactions parameters leads us to determine the phase diagrams of  the 1D Bose clouds affected by three different external fields; free homogeneous, harmonic, and optical lattice confining potentials. These results were used to propose long-life states composed of alternate assemblies of individual species confined in optical lattices that mimics magnetic domains, whose size can be adjusted. Our analysis based on numerical experiments within a mean field scheme considered large enough systems as those commonly used in experimental platforms.
\end{abstract}

\maketitle

\section{Introduction}
Originally, the unique way of adding inter-particle interactions to dilute ultracold gases was by manipulating the Zeeman splitting levels through an external magnetic field. As well known, such interactions besides being isotropic because of the fact that alkaline atoms are spherically symmetric in its ground state, are short ranged since the induced interactions have the form of a van der Waals potential type \cite{Pethick}. Although it is true that such short range interactions, also termed -contact interactions- have proven to be an essential ingredient to simulate a large variety  of quantum matter behavior, including dynamics, properties, as well as remarkable phase transitions \cite{Engelbrecht,Ohashi_PRL02,Giorgini-Stringari,Ketterle-review,Feshbach-review,Chiofalo,strinatibcs,Ohashi-review}, a most general picture of matter demands to include to go beyond, encompassing anisotropic long-range interactions from weak to strong. There are several schemes through which these kind of interactions can be harnessed. Among them, phonon-mediated interactions in trapped ion systems \cite{Blatt}, optical cavity or waveguide mediated interactions \cite{Ritsch, Mivehvar}, dipole-dipole interactions (DDI) by inducing electric dipole moments in heteronuclear molecules \cite{Carr, Moses, Bohn, Saffman,Low,Bernien}, and of course, magnetic DDI in magnetic atoms that offer long life times.

The first realization of a Bose-Einstein condensate with atoms having an exceptionally large magnetic dipole moment happened in $^{52}$Cr atoms in 2004 \cite{Pfau2005}. Besides being an hydrogen like atom with an electron in its outer shell, chromium atoms posses a magnetic moment as large as $\mu_0 = 6 \mu_B$, with $\mu_B$  the Bohr magneton. With no doubt, the capability of tuning DDI in magnetic systems opened the possibility of accessing the quantum matter regime combining both, short and long range interactions. In fact, diverse many-body phenomena, as for instance collapse, stability and structure formation strongly depends on competing contact vs. anisotropic dipolar interactions. Within the quantum phases accesible from competing weak contact against strong DDI, one can mention, droplet assemblies, self-bound droplets \cite{Ferrier, Ferrier1}, and supersolid regimes, being this last, matter having crystalline order and superfluidity at the same time \cite{Chomaz,Modugno, Pfau, Chomaz2,Stringari,Pohl}. It is imperative to point out that translational invariance in supersolid phases is broken not as a result of an external field as those proper of stationary patterns of light, but as a consequence of the long range DDI \cite{Lang}. 
\begin{figure}
\centering
\includegraphics[width=6.5cm]{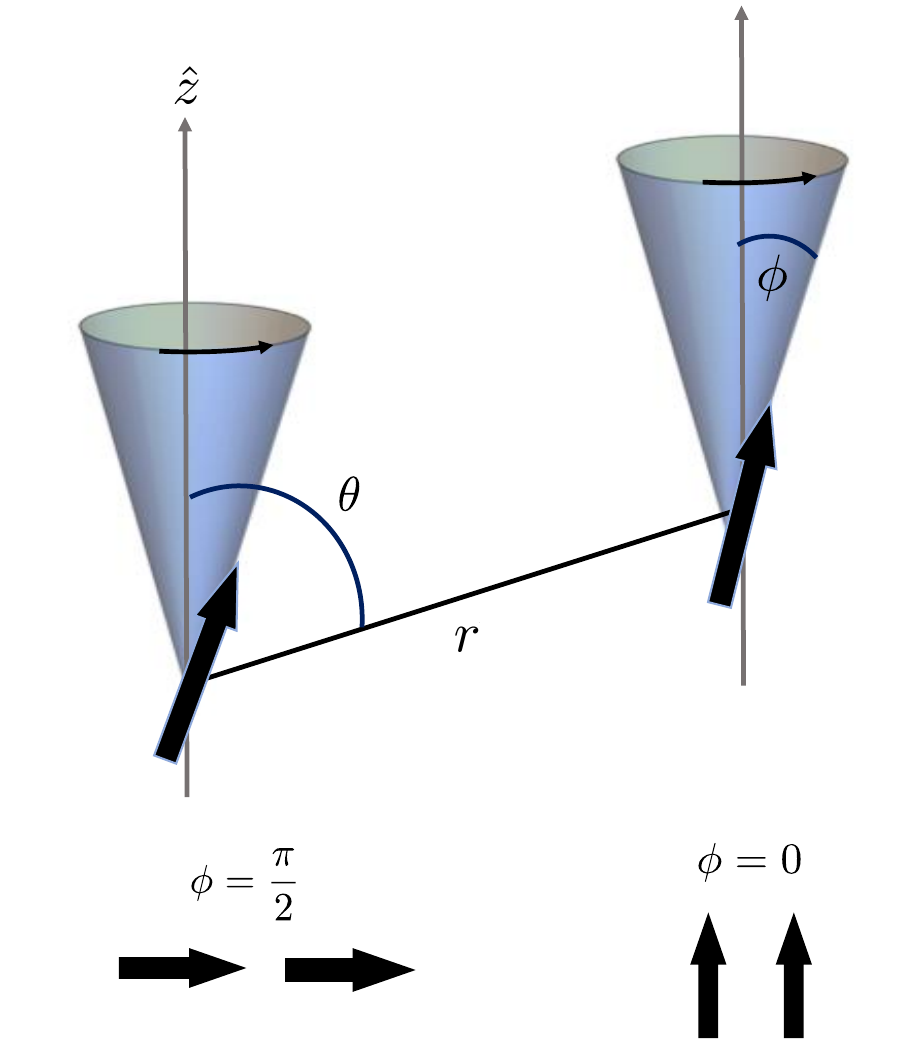}
\caption{A time-varying magnetic field gives the possibility of tuning the effective interaction among pairs. A quasi-one dimensional condensate composed of a doble species system.}
\label{Fig0}
\end{figure}
Besides the variety of interactions alluded in the above paragraphs, another type of energy present in ultracold atoms  arises from inhomogeneous external potentials resulting from static magnetic and electric fields. We refer to the typical harmonic trapping confinement and stationary patterns of light, the so called optical lattices, where the atoms develop its dynamics. Dimensionality and size can be controlled by properly adjusting parameters of the external fields. 

In this article we consider a binary mixture of Er and Dy clouds placed in three different external fields in 1D. The atoms in the clouds, having a large magnetic moment \cite{Lu,Lu2,Aikawa,Aikawa2}, are under the influence of a time dependent magnetic field that allows to vary the magnitude and sign between intra- and inter- DDI. Fig. \ref{Fig0} illustrates the configurations supporting attractive and repulsive interactions. Based on the knowledge of stable mixed, demixed, and droplet regimes that the binary mixture has, we propose a protocol for preparing long-life states that can be designed using realistic experimental parameters. These states consist of alternate sets of each component occupying the sites of an optical lattice, in such a way that a fixed number of sets can be prepared.

The manuscript is organized in five sections. Section II is dedicated to describe the model system, as well as the route to analyze the stationary regimes that appear in presence of both contact and DDI. In section III we focus our attention on the phase diagrams separating mixing-demixing of the erbium and dysprosium atoms. We propose a protocol to prepare states enclosing a fixed number of erbium and dysprosium atoms in section IV. Finally in section V we outline our findings and present some perspectives. 

\section{Two-component DBEC}
\label{Model}

Our system is composed of a double-species balanced mixture of Bose Einstein condensates, whose neutral atoms have a large magnetic moment. The Bose cloud has the form of a quasi-1D cigar that arise, as we describe below, from adjusting the harmonic frequencies along a certain axis. The composite system is under the presence of a rotating external magnetic field ${\vec B}({\vec r},t)= B \left [\cos {\phi} \>\hat{z} + \sin{\phi} \left ( \cos{\Omega t} \> \hat{x} + \sin{\Omega t} \> \hat{y} \right) \right]$, being the frequency $\Omega= 2 \pi \times 1$ kHz \cite{Tang}. This field produces effective DDI of two types, attractive or repulsive. In figure \ref{Fig0} we illustrate the configuration of two dipoles, belonging the 1D cloud in presence of the field ${\vec B}({\vec r},t)$. Since the cloud is in 1D, the angle $\theta$ that defines the dipole orientations separated a distance $\vec {r}$ is $\theta =\pi/2$. The angle $\phi$ defines the direction acquired by the dipoles under the presence of ${\vec B}({\vec r},t)$.

The components of the mixture, that is $^{166}$Er and $^{164}$Dy clouds occupying the same space, are described by the wave functions $\psi_1(x,t)$ and $\psi_2(x,t)$ respectively. Within the mean-field formalism these wave functions $\psi_1(x,t)$  and $\psi_2(x,t)$ satisfy the coupled dipolar Gross-Pitaevskii equations,

\small
\begin{equation}
\begin{split}
i\hbar\frac{\partial\psi_{1}}{\partial t}=&\left[H_{1}+C_{11}\left|\psi_{1}\right|^{2}+C_{12}\left|\psi_{2}\right|^{2} +D_{11}\frac{\partial^2}{\partial x^2}\left(U(x)*\left|\psi_{1}\right|^{2}\right)\right.\\
&\left.+D_{12}\frac{\partial^2}{\partial x^2}\left(U(x)*\left|\psi_{2}\right|^{2}\right)\right]\psi_{1}, \\
i\hbar\frac{\partial\psi_{2}}{\partial t}=&\left[H_{2}+C_{22}\left|\psi_{2}\right|^{2}+C_{21}\left|\psi_{1}\right|^{2} +D_{22}\frac{\partial^2}{\partial x^2}\left(U(x)*\left|\psi_{2}\right|^{2}\right)\right.\\
&\left.+D_{21}\frac{\partial^2}{\partial x^2}\left(U(x)*\left|\psi_{1}\right|^{2}\right)\right]\psi_{2},
\end{split}
\label{coupledGP}
\end{equation}
where $H_i= -\frac { \hbar^ 2 }{2m_i} \frac{\partial^2}{\partial x^2} +V_{\mathrm{ext}}\left( x \right)$, with $m_i$ being the mass of the $i$ component, and $i=1,2$ the label for erbium and dysprosium respectively. $V_{\mathrm{ext}}\left( x \right)$ is the external potential in 1D. In this paper we shall consider three different potentials $V_{\mathrm{ext}}(x)$, an optical lattice, a harmonic trap, and the free homogeneous space. In the case of the optical lattice $V_{\mathrm{ext}}(x)= V_0 \cos^2\left( \frac{\pi x}{a} \right)$, being $V_0$ the potential deep, which is scaled in units of the recoil energy for the erbium atom $E_R=\frac{\hbar^2k^2}{2m_1}$, being $k=\frac{\pi}{a}$, with $a$ the lattice constant. The harmonic confinement is described by $V_{\mathrm{ext}}(x)= \frac{1}{2}m_i \omega^2 x^2$. Finally, the  box with infinite walls representing the homogeneous potential has a length $ \sim 40 a$. The contact interactions $C_{ij}$ are written in terms of the s-wave scattering lengths $a_{ij}$. The dipolar interaction couplings  $D_{ij}$ are written in terms of the individual dipolar magnetic moments $\mu_{i}$. These constants are given by the next expressions: 
\begin{equation}
\begin{split}
&C_{ij}=\frac{N_{j}}{2\pi l_{\perp}^{2}}\left[g_{ij}^{c}+\frac{g_{ij} ^{d}}{6}\right], \\
&D_{ij}=\frac{g_{ij}^{d}N_{j}}{8\sqrt{2\pi}l_{\perp}},\\
&g_{ij}^{c}=\frac{2\pi\hbar^{2}a_{ij}}{\mu},\\
&g_{ij}^{d}=\mu_{0}\mu_{i}\mu_{j}T,\\
&T=\frac{3 cos^{2}\phi-1}{2},
\end{split}
\label{Constantes}
\end{equation}
where $\mu= m_1 m_2/(m_1+m_2)$, $N_j$ is the atom number of $j$ component, in this paper we consider a balanced mixture of species, that is, $N_1=N_2=N$. The transverse harmonic length is $l_\perp = \sqrt{\hbar/m_1\omega_\perp}$, being $\omega_\perp$ a typical frequency used to confine condensates in 1D \cite{Hadzibabic, Hung}, and $\mu_0$ is the magnetic permeability of the vacuum. Depending on the value of the angle $\phi$, which is the angle that the polarization vector makes with the $z$ axis, $T$ takes values in the interval $[-0.5, 1]$. Also it should be noted that the $C_{ij}$ and $D_{ij}$ coefficients have already been rescaled to work with a one-dimensional equation \cite{Bao}. Finally $U(x)=e^{\frac{1}{2l_{\perp}^{2}}x^{2}}erfc\left(\frac{\left|x\right|}{\sqrt{2}l_{\perp}}\right)$  is the effective dipole interaction potential in 1D.

In this paper we shall focus on the role played by the long range DDI in boosting the opposite phase mixed and demixed states where either droplet or extended states arise. While the former is characterized by having both components situated in the whole space, in the latter individual components occupy definite spatial regions. In typical experiments with ultracold atoms the values of the s-wave scattering lengths $a_{ij}$ can be tuned through Feshbach resonances, thus modifying the effective contact interactions  $g_{ij}^{c}$, in such a way that the occurrence of mixing of the gas components can be controlled. If the contact interaction constants satisfy $g_{12}^{c}<\sqrt{g_{11}^{c}g_{22}^{c}}$ then the gas components will be miscible, otherwise the gas will be in a demixed or phase separated regime \cite{Wang}. As stated above, dipole-dipole interactions can be varied by means of a rotating magnetic field as follows $g_{ij}^{d}=\mu_{0}\mu_{i}\mu_{j}T$ \cite{Giovanazzi, Tang}.

\section{Mixture and phase separation}
To quantify the influence of dipole-dipole interactions on the miscibility and phase separation regimes of Er and Dy, as well as the formation of quantum droplets, we determine the stationary state of Eqs. \ref{coupledGP}. The imaginary time evolution method \cite{Muruganandam, Vudragovic, Loncar, Li}, which strongly depends on the initial seed, is used to find the double-species condensate stationary state. For generality we use an initial state having random values for both, real and imaginary components of $\psi_1^0(x)$ and $\psi_2^0(x)$. Importantly, beyond mean-field effects like the Lee-Huang-Yang (LHY) correction are not needed to reach stability for the droplet state \cite{LHY}.

\begin{figure}
\centering
\includegraphics[width=8.7cm]{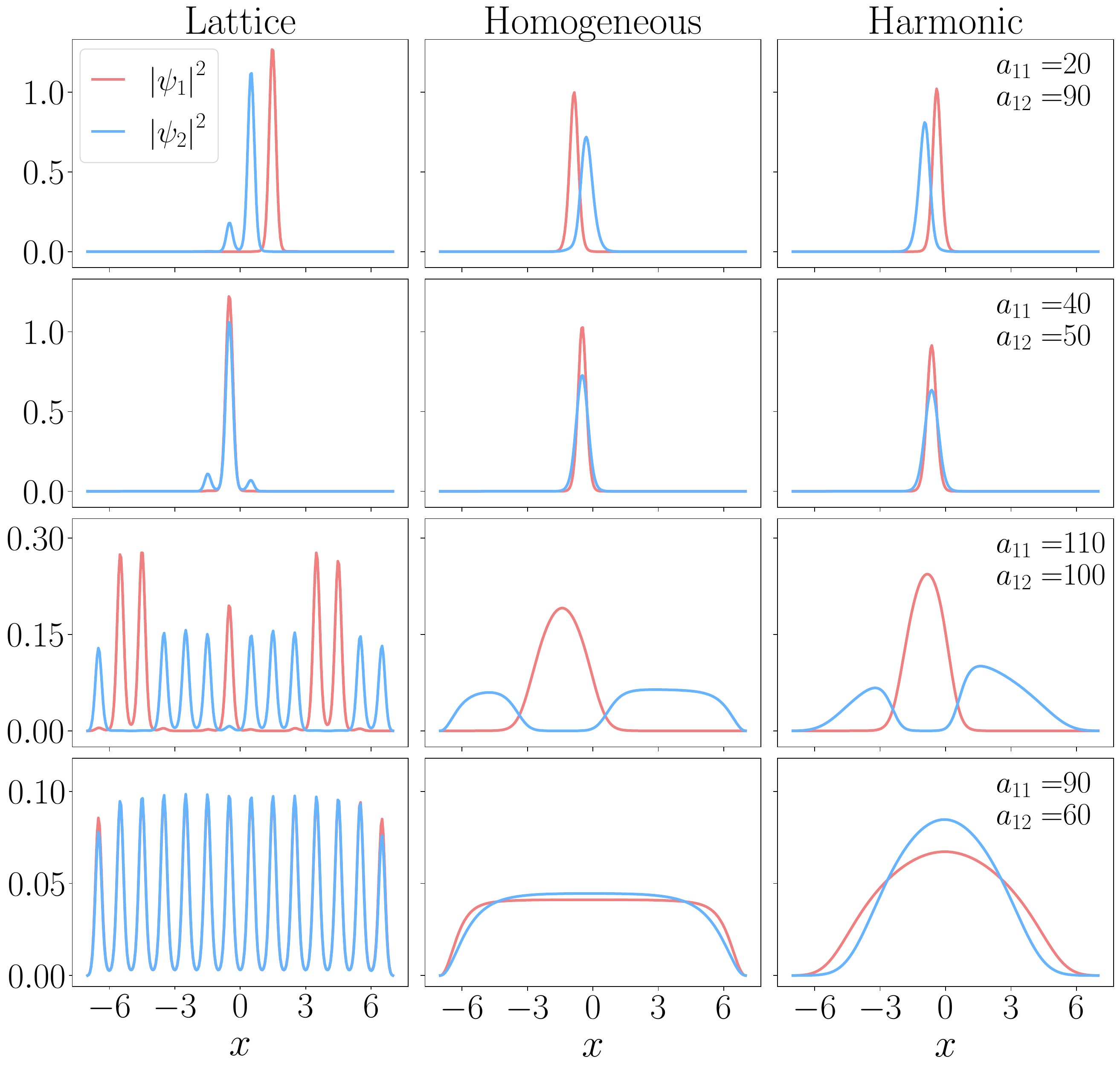}
\caption{Stationary states of the system. The columns from left to right indicate the external potential considered, that is, Lattice, Homogeneous and Harmonic trap. Different rows indicate the possible states: from top to bottom we have separated droplets, mixed droplets, demixed extended states, and spread mixtures.}
\label{Fig_1}
\end{figure}

Depending on the coupling interaction parameters $a_{ij}$ and $T$, different configurations can be found for the stationary regime. In figure \ref{Fig_1} we illustrate the possible states that can be obtained. While for the first and second rows $T=-1/2$ (that is $\phi=\pi/2$), third and fourth rows correspond to $T=1$. For the effective Dy-Dy contact interaction, we chose a typical value $a_{22}=92 a_B$, while the specific values of $a_{11}$, and $a_{12}$ are indicated on the right of the third column. As expected, one can appreciate four different possibilities: separated droplets (first row), mixed droplets, that is droplets placed in the same region (second row), separated clouds (third row) and mixed clouds, namely, clouds filling the whole space (fourth row). One can also appreciate the influence of the external potential $V_{\mathrm {ext}}(x)$ from the density profiles. Left, center and right columns correspond to lattice, homogeneous, and harmonic confinements respectively. 

\begin{figure}[H]
\centering
\includegraphics[width=7.0cm]{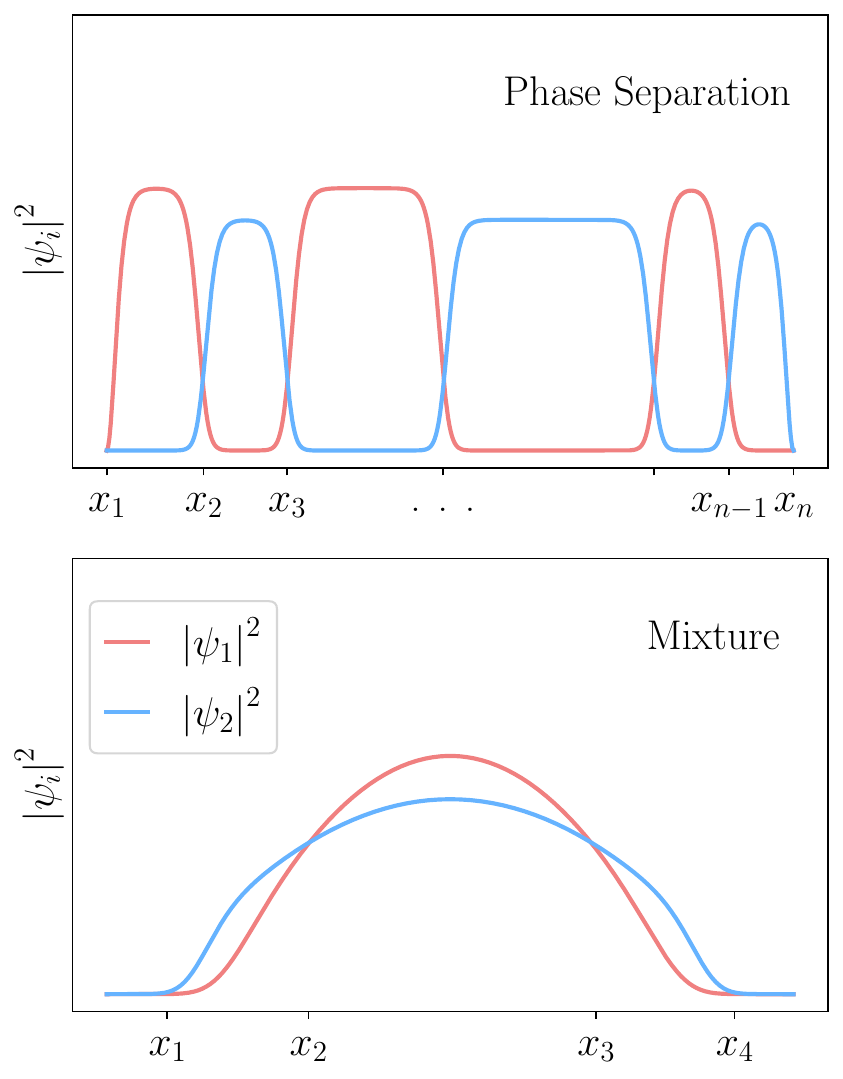}
\caption{Fragments of the density profiles associated with the stationary states for the double-species Er-Dy condensate system. Top and down figures represent phase separation and mixed states respectively.}
\label{Fig_0}
\end{figure}

To construct a phase diagram with the possible phases that emerge as a function of the coupling interaction parameters, it is useful to introduce the following quantities since they capture the correspondence among a given density distribution of the Er and Dy and its associated phases. First, for the miscible and inmiscible regimes we define $PS$ as follows,
\begin{equation}
PS(\psi_{1},\psi_{2})=\left|\sum_{j=1}^{n}(-1)^j  \int_{\Omega_{j}}\left[\left|\psi_{1}\right|^{2}-\left|\psi_{2}\right|^{2}\right]dx\right|
\label{Observable}
\end{equation}
where $\Omega_{j}$ is a region bounding the zone $x \in [x_j,x_{j+1}]$, such that at the points $x_j$ and $x_{j+1}$ both, the difference $\left|\psi_{1}\right|^{2}-\left|\psi_{2}\right|^{2}$ and the probabilities $\left|\psi_{1}\right|^{2}$ and $\left|\psi_{2}\right|^{2}$ become negligible. In Fig. \ref{Fig_0} we show a fragment of the combined densities $\left|\psi_{1}\right|^{2}$ and $\left|\psi_{2}\right|^{2}$ to illustrate the region $\Omega_i$ for both cases; when the components do not occupy the same space, thus being in a demixed regime, or when the components are placed in the same region. Notice that positions $x_i$ do not necessarily match the lattice positions. One can observe from the definition of $PS$ is that if the binary system is in a mixed (demixed) regime, then $PS$ will take 0 (1) value. As one can see from Fig. \ref{Fig_1}, no matters the nature of the external potential, blue and orange regions encloses condensate densities of Er and Dy species distributed in lattice or continuos space.

The quantity $PS$ allows us to detect the mixed and demixed regimes of the composite Er-Dy system. However, to discern the droplet regime, that is the sector of parameters for which densities of both species are tightly confined, either in the lattice or continuous space, another quantity must be used. We introduce the quantity $D(\psi_{1},\psi_{2})$ defined as follows,

\begin{equation}
\begin{split}
D(\psi_{1},\psi_{2})=\begin{cases}
1 \>\>\> { \mathrm{if}} \>\>\> |\psi_{1}|^2 \>\>\>{\mathrm{and}} \>\>\> |\psi_{2}|^2   = 0  \>\>\>   \forall \>x  \in I.\\
0 \>\>\> { \mathrm{if}} \>\>\> |\psi_{1}|^2 \>\>\> {\mathrm{or}} \>\>\> |\psi_{2}|^2   \neq 0 \>\>\>   \forall \>x \in I.
\end{cases}
\end{split}
\label{Ext_pot}
\end{equation}
that is, to construct a phase diagram for the droplet regime, we assign values of 1 or 0 to $D(\psi_{1},\psi_{2})$ according with the following criterium. If for all $x$ belonging to $I$, being $I$ about  $\sim$95\% of the whole space, and the densities of each species are nearly zero ($|\psi_i(x)|^2 \approx 0$), we assign 1. Conversely, if the densities are non negligible, also in the region $I$, we assign 0. Then, we see that the quantity $D(\psi_{1},\psi_{2})$ simply registers if the state associated to certain coupling interaction parameters is an extended state or belongs to a droplet regime.

The phase diagrams enclosing the whole states of the bi-component system were constructed selecting interaction parameters $a_{ij}$ that can be accessed experimentally for Er and Dy species \cite{Chomaz, Ferlaino, Ferlaino2, Adhikari}. The protocol followed in our numerical simulations was the following. As explained before, random seeds were considered for real and imaginary components of wave functions $\psi_1(x,t)$  and $\psi_2(x,t)$. Subsequently, evaluation of quantities $PS$ and $D$ in the stationary states $|\psi_1(x)|^2=|\psi_1|^2 $ and $|\psi_2(x)|^2=|\psi_2|^2$, leads us to classify the phase associated to a given set $T$, $a_{11}$, $a_{22}$ and $a_{12}$. Thus, the corresponding phases for the whole parameter space explored were obtained.  All the results presented below correspond to a spatial window of 40 lattice sites. In order to avoid finite size effects in our numerical calculations we tested for spatial regions as large as 320 lattice constants for both homogeneous and harmonic cases. The results found for the frontiers separating mixture, droplet, and phase separated regimes were the same as those associated to 40 lattice sites. 
  
\begin{figure*}
  \centering
   \includegraphics[keepaspectratio=true,scale=0.315]{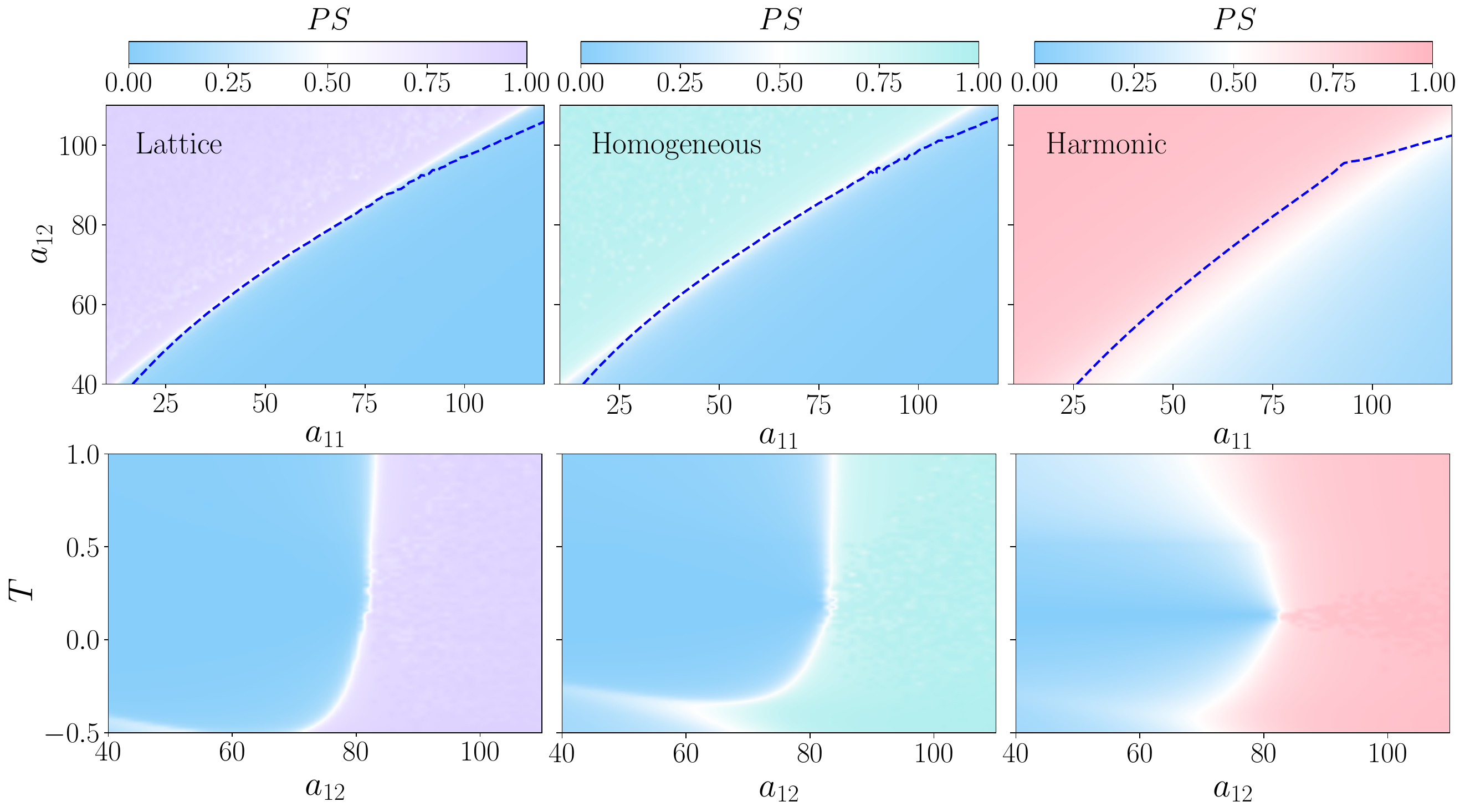} 
  \caption{The quantity $PS$ defined in \ref{Observable} is plotted in a color scale. In the upper panels, $a_{12}$ and $a_{11}$ were varied, keeping $T=1$ and $a_{12}=92a_B$. The blue line indicates the boundary between the mixing and separation regions when $T=0$. In the lower panels, $T$ and $a_{12}$  were varied, keeping $a_{11}=72a_B$ and $a_{22}=92a_B$. The different colors distinguish the type of potential used: optical lattice (purple), homogeneous (cian), and trap (pink). In these figures, the blue region represents the system in a mixed state, while in the other areas, the system is in a separated state.}
  \label{Fig_2}
\end{figure*}

\begin{figure*}
  \centering
  \includegraphics[keepaspectratio=true,scale=0.315]{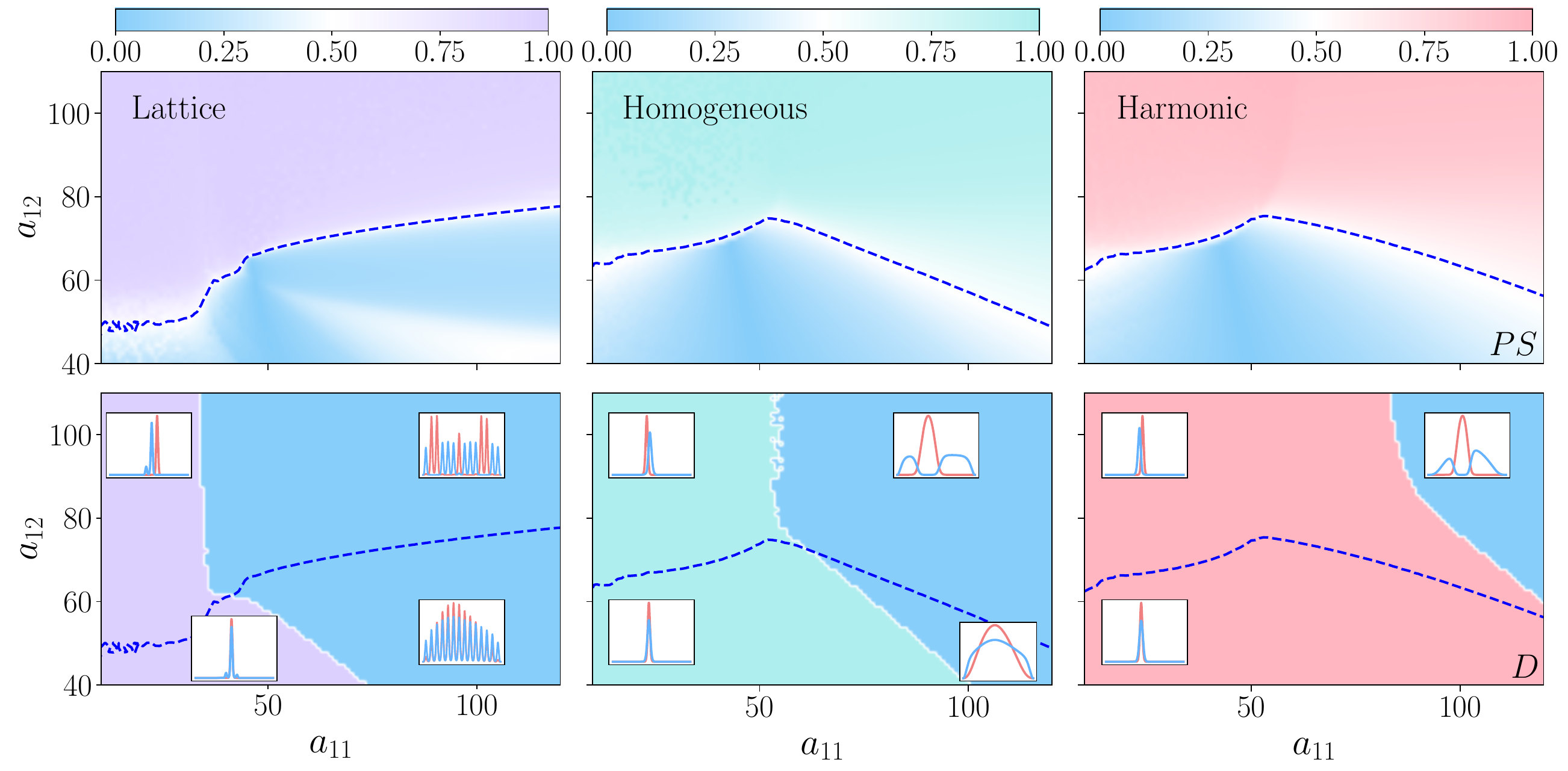} 
  \caption{In the upper and lower panels, the observables $PS$ and $D$ are plotted in a density color scale, respectively. In both, $a_{12}$  and $a_{11}$  are varied while maintaining $T=-0.5$ and $a_{12}=92a_B$. The different colors distinguish the type of potential used: optical lattice (purple), homogeneous (cian), and trap (pink). In the upper panels, the blue regions represent mixing states, while the regions with other colors represent separation states. In the lower panels, blue regions indicate extended states, while the other colors indicate droplet states. The blue dotted line is the border that divides the separation and mixing zones. Thus, in the lower panels, there are four different regions corresponding to the states: separated droplets, mixed droplets, separated extended states, and mixed extended states. These four possible states are illustrated in the insets of each figure.}
  \label{Fig_3}
\end{figure*}

Figure \ref{Fig_2} summarizes in a density color scheme the results found for mixed and demixed phases of the double component condensate placed in the three different external fields: lattice, homogeneous, and harmonic confinements. While in the upper panels one observes the opposite mixed (blue) and demixed (purple, cian and pink colors) regimes as a function of $a_{12}$ and $a_{11}$ for constant values of $T$ and $a_{22}$ ($T=1$, $a_{22}= 92a_B$), in the lower panels the corresponding phases (same color scale) are diagnosed in terms of $T$ and $a_{12}$ keeping $a_{11}=72 a_B$ and $a_{22}={92 a_B}$. As already known \cite{Wang} and mentioned above, miscibility occurs in presence of contact interactions when the $s-$wave scattering lengths satisfies the condition $a_{12} < \sqrt{a_{11} a_{22}}$. It is expected that such a tunable miscibility becomes affected by is the presence of DDI. Blue dashed curve in the upper panels indicates the boundary that divides mixed and demixed phases in the absence of DDI, that is when $T=0$ is considered. This frontier is defined as the region for which $PS=0.5$. One can see that such a boundary is very similar to the case $T=1$ (white region associated with $PS=0.5$) for the lattice and homogeneous potentials. A smooth transition is appreciated for the harmonic confinement as DDI are non-zero. In this case, repulsive interactions compete against  harmonic trapping, boosting the phase separation between components, and thus stretching the blue region in the phase diagram.

The behavior of $PS$ as a function of $T$ and $a_{12}$, and keeping $a_{11}=72a_B$ and $a_{22}=92a_B$ fixed is illustrated in lower panels of figure \ref{Fig_2}. As expected for $T=0$ and $a_{12}> 81.38 a_B = \sqrt{a_{11} a_{22}}$, the binary system enters into the phase separation regime. However, when $T$ takes positive values one does not anticipate significative changes in the frontier for the lattice and homogeneous confinements since the effective contact interactions $c_{ij}$ maintain an analogous inequality satisfied by the nude contact interactions, that is $C_{12}< \sqrt{C_{11} C_{22}}$. What happens for the harmonic trapping is that at the center of the trap DDI are magnified since the density is larger as compared with the other external potentials. When $T<0$ the value of the effective contact interactions $C_{ij}$ decreases, thus promoting the phase separation regime. This occurs because around the frontier one obtains $C_{12}> \sqrt{C_{11} C_{22}}$.

On the other hand, having $T<0$ makes dipolar interactions attractive, which favors the formation of quantum droplets. For the values considered for $a_{11}$ and $a_{22}$, the area where quantum droplets form is for small $a_{12}$  and $T<-0.4$. This can be explained by the fact that droplet formation occurs when the long-range attractive interactions between the components of the condensate compete with the repulsive contact interactions. Thus, having large scattering lengths $a_{ij}$  prevents the formation of droplets, whereas if the contact interaction is small, droplet formation becomes possible.

To investigate the droplet formation regime, we first determine the phase diagram for mixed-demixed states in the attractive DDI sector. For this purpose we consider the same set of contact interaction parameters than those employed in fig. \ref{Fig_2} (upper panels), but instead of considering repulsive DDI, we set $T=-0.5$. The results summarized in fig. \ref{Fig_3} (upper panels) indicate  very different phase diagrams with respect to those obtained for $T = 0$. The color bar on the top allows to identify the mixing-demixing regions. Again, blue dashed curve is the frontier associated to $T=0$. We observe demixed dominated phase diagrams, disregarding the presence of the external potential. Now with the use of the quantity $D$ we examine the space of parameters where droplets are formed. Lower panels of fig. \ref{Fig_3} show in a density color scheme the sectors where different types droplets arise when DDI are attractive. Purple, cian and pink regions correspond to values of $a_{11}$ and $a_{12}$ for which droplets are formed. However, two types of droplets can be detected, droplets occupying the same space or droplets situated at different places. These two possibilities depend on the values of the contact interaction parameters. If the triad $(a_{11}, a_{22}, {a_{12}})$ is such that the binary system belongs to a demixed phase, then the droplets of erbium and dysprosium will be placed in different places, on the contrary, if the triad corresponds to the mixed species phase, then droplets will occupy the same spatial point. In lower panels of fig. \ref{Fig_3} insets with densities of erbium and dysprosium illustrate the types of quantum droplets associated to each external potential.

\section{Protocol for DBEC domains}
Upon examining the stationary states exhibiting phase separation, and in particular those in which the binary mixture is placed in an optical lattice, we performed a systematic analysis to track for the resulting states that can be found in a superposition of both, a harmonic potential plus an optical lattice confinement. This combination is a realistic one since typical experiments involving ultracold dipolar gases in optical lattices are always in presence of an inhomogeneous environment \cite{Ferrier, Ferrier1}. Based on such analysis, we propose here a protocol were robust states with specific attributes can be prepared. These states consist of alternated domains or regions where the DBEC's are located. We must stress here that typical values of atoms are used to design the configurations of alternate single components. As we shall see, the number of such domains crucially depends on the specific values of the contact and DDI coupling parameters. 

\begin{figure}
  \centering
  \includegraphics[width=0.42\textwidth]{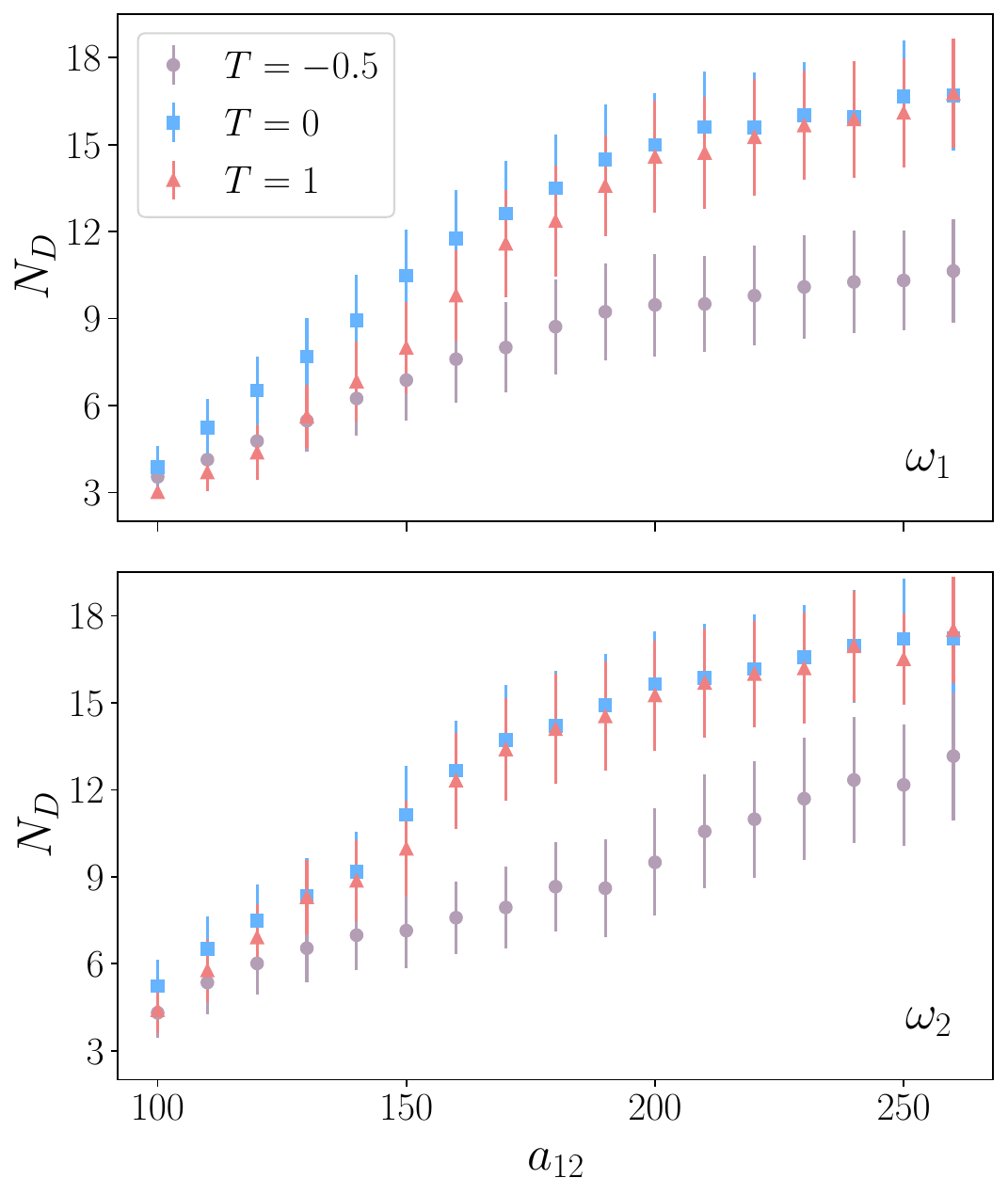} 
  \caption{The number of regions is plotted as a function of the inter-species $s$-wave scattering length $a_{12}$, we consider $a_{11}=72a_B$ and $a_{22}=92a_B$. Top and bottom panels correspond to different values of the harmonic confinement ($\omega_1>\omega_2$). Symbols in different colors label the values of the parameter $T$, which controls the dipolar interaction.}
  \label{Fig_4}
\end{figure}

\begin{figure}
  \centering
  \includegraphics[width=0.42\textwidth]{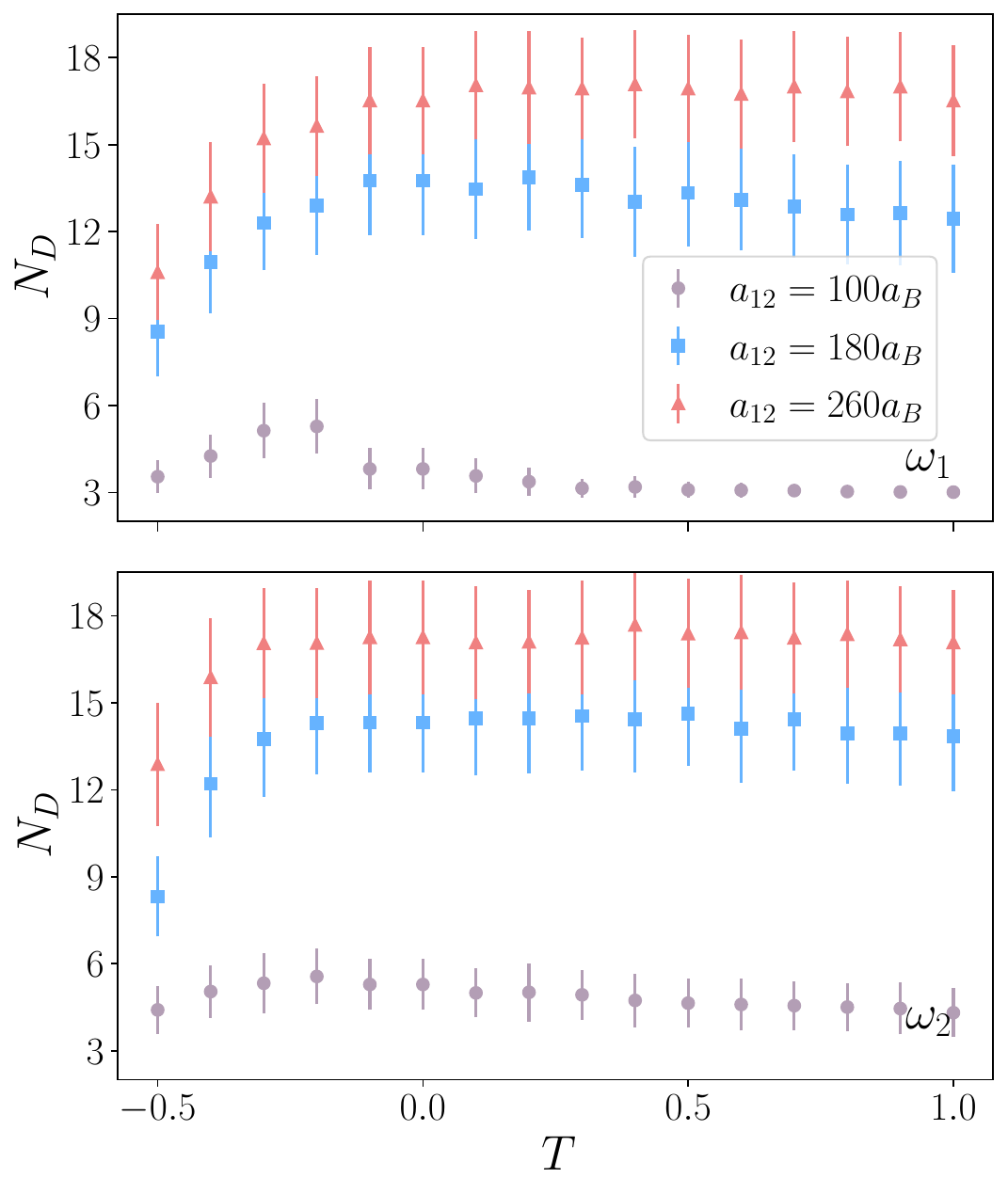} 
  \caption{The number of regions is plotted as a function of the parameter $T$,  we consider $a_{11}=72a_B$ and $a_{22}=92a_B$.. Top and bottom panels correspond to different values of the harmonic confinement ($\omega_1>\omega_2$). Different symbols identify different values of the $s$-wave scattering length, which controls the contact interaction.}
  \label{Fig_5}
\end{figure}

In previous section we found that in presence of a harmonic confinement, the phase separation regime is such that the DBEC components are situated occupying three regions. For $T \geq 0$, two clouds of erbium are always surrounding a cloud of dysprosium. However, when $T$ takes negative values, two possibilities can occur, droplets or clouds can be found depending on the values of the triad $a_{12}$, $a_{11}$ and $a_{22}$. Whereas in the droplet regime the binary mixture is settled in two prominent droplets, therefore, the number of regions is equal to two.

On the other hand, in an optical lattice, the number of regions is unstable, unless a harmonic confinement is superimposed to the lattice array. In other words, we must use the harmonic confinement to stabilize the formation of regions in an optical lattice. To achieve this, we use contact interaction coefficients $a_{ij}$ that ensures the maintenance of a cloud regime. In particular, we analyze how the number of regions formed behaves as a function of $T$ and $a_{12}$. It is important to stress here that the values of $a_{12}$ used in the analysis correspond to reliable experimental ones. Also, we must stress that our study included two different values of the harmonic frequency $\omega$ ($\omega_1>\omega_2$) to show the negligible influence on it.  

Figures \ref{Fig_4} and \ref{Fig_5} summarize the results of the analysis described in previous paragraphs. $N_D$ is the number of domains or regions having a definite component $\psi_{i}$ placed on the lattice sites. The number of individual lattice sites in each domain is equal to $ \approx 40/N_D$. As can be seen from figures \ref{Fig_4} and \ref{Fig_5}, the number of alternated domains, $N_D$, depends on both $a_{12}$ and $T$. For attractive dipolar interactions $N_D$ is quite different against the values it takes for repulsive DDI, specially for values of $a_{12}  \sim \geq 200 a_B$. As can be appreciated from fig.\ref{Fig_4}, for $a_{12} \leq 150$, $N_{D}$ grows linearly with a certain slope, and then grows with other rate for $a_{12}  \geq  200 a_B$. Fig. \ref{Fig_5} illustrates how stable the values of $N_D$ are as $T>0$. Interestingly, for values of $a_{12} \sim 100 a_B$ a reduced number of domains is found ($\sim 3$). In contrast, for larger values of $a_{12}$, $N_D \sim [15,18]$. It is worth to mention here that previously efficient design of quantum protocols in a one-dimensional optical lattice have been reported \cite{Dupont, Haber}.  

The procedure described above allows us to recognize the correspondency among a given state and the particular values of the tetra $(a_{11}, a_{22}, a_{12}, T)$. In other words, we hypothesize that each state prepared is a single valued one. Therefore, we conclude that for fixed values of the contact interaction $s$-wave scattering length parameters $a_{ij}$, a particular configuration of the binary mixture, can be associated to a specific magnitude of the magnetic field $B=|{\vec B}({\vec r},t)|$. In this way, we have proposed here a technique for initializing quantum states, and at the same time, a procedure to read a given quantum state by measuring a magnetic field. We should point out that knowledge of the quantities $PS$ and $D$ is not essential, and therefore sensitivity of the state is not imperative.

\section{Conclusions}
\label{Conclusion}
A binary mixture composed of dipolar Bose-Einstein condensates of $^{168}$Er-$^{164}$Dy placed in one dimensional free, harmonic, and lattice confinement potentials was investigated in this manuscript. The clouds nested in an external magnetic field ranged dipole-dipole interactions from repulsive to attractive, thus competing with the intrinsic weak contact interactions of the species. 

Our analysis done in the mean-field frame considering systems with a size comparable to those in its experimental counterpart allowed for a full exploration of the parameter space composed of inter- and intra-species, and dipole-dipole interaction couplings. Two quantities defined in this work were evaluated in the stationary state, thus allowing us to construct the mixing-demixing phase diagrams, as well as end up with a classification of types of droplets that emerge in the 1D system. These results aimed us to propose a protocol to be set in the current platforms of ultracold atoms. The robust stationary states consist of spatial regions where alternate fringes of each specie remain occupying the sites of an optical lattice. These long lived states can serve either, as information storage devices, or for quantum computing purposes.

Future perspectives of the present study include two-dimensional dipolar Bose-Einstein condensates where more complex states resulting from large and anisotropic interactions can emerge as for instance,   droplet assemblies and supersolid phases. Also, multi-component Bose Einstein condensates (spinor-BEC's) in two and three dimensions to get a deep understanding of magnetic domains dynamics and non-linear excitations between others.  

\acknowledgments{
This work was partially funded by Grant No. IN117623 from DGAPA (UNAM). C.J.M.C acknowledges CONACYT scholarship.
}

\appendix

%\section{A}
%\label{AppendixA}

%\appendix
\section{Numerical Calculations Details}
\label{AppendixB}
The results here reported were obtained from numerical solution of the coupled GP equations. In this appendix we present additional details related to the technical part of the calculations performed. To find the solution of the Eq. \ref{coupledGP} we use the fourth-order Runge-Kutta method, to obtain the stationary of the system. The numerical parameters used for the simulations are shown in table \ref{Table1}, while the physical parameters of the system are presented in table \ref{Table2}.

\begin{table}[H]
\centering
%\begin{center}
\small
\begin{tabular}{p{4.5cm} c  c} 
Name & Symbol & Value\\
\hline\hline
%\hline
Number of grid points in the $x$ direction & $N_{x}$ & 512-4096\\
%\hline
Spatial extension of the numerical grid in the $x$ direction & $L_{x}$ & 40-320 $a_0$ \\
%\hline
Step size used in real time evolution & $d\tau$ & $0.0005$\\
\end{tabular}
%\end{center}
\caption{Parameters for the numerical simulation}
\label{Table1}
\end{table}

\begin{table}[h]
\centering
%\begin{center}
\small
\begin{tabular}{p{4.5cm} c  c} 
Name & Symbol & Value\\
\hline\hline
Particle number & $N$ & 300\\
$^{166}$Er mass & $m$ & $165.9$ amu\\
%\hline
$^{164}$Dy mass & $m$ & $163.9$ amu\\
%\hline
$s$-wave scattering length between Er, Er & $a_{11}$ & 72 $a_{B}$ \\
$s$-wave scattering length between Dy, Dy & $a_{22}$ & 92 $a_{B}$ \\
$s$-wave scattering length between Er, Dy & $a_{12}$ & [40, 267] $a_{B}$ \\
%\hline
dipolar magnetic moment for Er & $\mu_{1}$ & 7 $a_{B}$ \\
dipolar magnetic moment for Dy & $\mu_{2}$ & 10 $a_{B}$ \\
%\hline
Lattice constant & $a$ & $532$ nm\\
%\hline 
Reference trap frequency & $\omega_{0}$ & $2\pi\times 50$ rad/s\\
Trap frequency $(z)$ & $\omega_{z}$ & $2\pi\times 5000$ rad/s\\
%\hline
Potential depth  & $V_{0}$ & 4 $E_R$  \\
\end{tabular}
%\end{center}
\caption{Physical parameters used in the numerical simulation}
\label{Table2}
\end{table}

\end{document}